\documentclass[a4paper]{jpconf}
\usepackage{graphicx}
\usepackage[export]{adjustbox}
\usepackage{url}
\usepackage{mcite}
\usepackage{hyperref}
\begin{document}
\title{CERNLIB status}

\author{Ulrich Schwickerath$^1$, Andrii Verbytskyi$^2$}
\address{${1}$~CERN, 1211 Meyrin, Switzerland}
\address{${2}$~Max-Planck-Institut f\"ur Physik, 80805 Munich, Germany.}
\ead{Ulrich.Schwickerath@cern.ch}

\begin{abstract}

We present a revived version of CERNLIB, the basis for software
ecosystems of most of the pre-LHC HEP experiments. The efforts to
consolidate CERNLIB are part of the activities of the Data Preservation
for High Energy Physics collaboration to preserve data and software of
the past HEP experiments.

The presented version is based on CERNLIB version 2006 with numerous
patches made for compatibility with modern compilers and operating systems.
The code is available in the CERN GitLab repository with all
the development history starting from the early 1990s. The updates also
include a re-implementation of the build system in CMake to ensure CERNLIB
compliance with the current best practices and to increase the chances of
preserving the code in a compilable state for the decades to come.

The revived CERNLIB project also includes updated documentation, which we
believe is a cornerstone for any preserved software depending on it.

\end{abstract}

\section{Introduction}
CERNLIB~\cite{Brun:cernlibshortwriteups} is a set of C and Fortran libraries developed  between the 1970s and 2000s. It was and is widely utilized in many high energy physics (HEP), nuclear physics, and astro-particle physics experiments for data analysis for almost four decades. 
As a result, the experiment-specific software for many old experiments and  even the access to the collected data depends on the availability of a functioning CERNLIB.
For instance, the  LEP experiments~\cite{Callot:623411} ALEPH, DELPHI, L3, OPAL, as well as non-CERN based experiments, such as JADE~\cite{Bethke:2022cfc}, heavily depend on CERNLIB.

The development of the CERNLIB, historically coordinated by the CERN IT department, ended officially in 2006. Since that time, the official sources, released under GPL license (with exceptions), were no longer updated. As a consequence, the code was picked up by numerous groups and communities around the globe, and further patches and adaptations to new platforms and compilers were done in an uncoordinated manner.

Given the importance of the CERNLIB for the old experiments, the DPHEP~\cite{DPHEPStudyGroup:2012dsv} collaboration has planned 
and successfully executed a project to consolidate the CERNLIB developments that happened after 2006 to assure a safe future of CERNLIB. 

\section{CERNLIB status before 2022}

The code base of CERNLIB contains 1600 kLOCs of Fortran77, 
and 500 kLOCs of C89. The code written in C is used primarily to provide an interface to system libraries. 
CERNLIB is organized in sub-packages according to their functionality.
Briefly, the  CERNLIB sub-packages included  basic utility subroutines and functions ({\sc kernlib}),  mathematical routines and random number generators ({\sc mathlib}), the graphics libraries({\sc graflib}), physics analysis tools such as the PAW~\cite{Bock:1987mf} library ({\sc pawlib}), as well as incorporated software. Namely, the Monte-Carlo event generators (MCEGs) such as Pythia6~\cite{Sjostrand:2006za} or ARIADNE~\cite{Lonnblad:1992tz} were included in the  {\sc mclib} sub-package. The GEANT3~\cite{Brun:1987ma} simulation toolkit with a G-FLUKA~\cite{Fasso:1993kr} interface were included as {\sc geant321} sub-package.

 The original build system of CERNLIB consisted of imake~\cite{imake} configuration files and shell scripts.
 
 The last release of CERNLIB was officially supported on a dozen of different Unix/Unix-like platforms and Windows.
Older versions of CERNLIB were also supported on many other platforms, e.g.\ VMS.
However, the level of CERNLIB support on these platforms varied and some sub-packages had only limited support for 64-bit platforms.

 Therefore, at the end of the 2010s, when the 64-bit platforms started to dominate HEP computing, numerous efforts were made to allow the usage CERNLIB on the most popular platform in HEP -- Linux x86\_64.

 After 2006 the CERNLIB builds were maintained by the Fedora (till 2016) and Debian projects.
These projects also collected hundreds of patches for CERNLIB.
A lot of those patches originated from the activities of Kevin McCarty to package the CERNLIB for Debian Linux. The second significant
group of patches was created by Harald Vogt and his collaborators from the DESY-Zeuthen group. In addition, input from private builds by the DELPHI collaboration was considered.
All those changes became the basis for the consolidation of CERNLIB in 2022.

\section{CERNLIB consolidation timeline}

The principal decision to consolidate and revive CERNLIB was taken in December 2021.
 In January of 2022, the collection of all information on the CERNLIB patch-sets and forks started. 
 In parallel, the original CVS repositories (one per sub-package)  were recovered and converted into git repositories. The resulting git repositories were put into \url{https://gitlab.cern.ch}.
Later, the separate repositories were merged into one git repository, and the build scripts were added.

 Between February and June 2022 the collected patch sets were carefully reviewed, broken down into smaller updates where needed, and added one by one to the code base. 

With a sufficient number of patches it became possible to 
compile CERNLIB on modern and historical Linux systems on x86\_64 architecture with the GCC compilers.
 To manage the increasing number of supported systems a continuous integration system (CI) based on GitLab CI  was introduced. 
 The development was focused on the Linux i686 and Linux x86\_64 platforms with GCC at this time, as only those systems had solid support in the CERNLIB native build system.
 
 In June of 2022 the CERNLIB build system, based on imake, with a set of build scripts was complemented by an alternative CMake based build system. 
 The comparison between the builds performed with imake and CMake setups was done using the compilation databases produced by the bear\cite{bear} tool on Linux systems. It was also checked that the resulting installations from both builds produce the same set of files.
 
The adoption of CMake has allowed support for a wider range of compilers and target systems to be explored. 
CERNLIB was compiled with the Intel One API compilers  on Linux x86\_64, NVidia HPC compilers on Linux x86\_64, GNU compilers on Mac OSX x86\_64,  GNU compilers on Solaris x86\_64, and GNU compilers  on Linux aarch64. The variety of compilers was beneficial for a more efficient elimination of bugs and incompatibilities with modern operating systems.

Further developments were related to the improvements in the CI, which tests the CERNLIB builds starting from very old systems (Scientific Linux 3) to very modern (Fedora 37). After gradual improvements and a certain amount of bug fixes, the updated CERNLIB was used for tests with the DELPHI software, by using it for a first proof of concept to run the entire DELPHI software stack in 64-bit mode on Linux.

After the preliminary tests, the first updated version of CERNLIB has been made available on the official CERNLIB site in November 2022 at 
\url{https://cernlib.web.cern.ch/cernlib/}. 
Along with this, the CERNLIB versioning schema was changed to a chronological one, with the date of the CERNLIB release in the name including release day, month and year.

After the release, the updated CERNLIB was added to the repositories of ARCHLinux, Homebrew~\cite{homebrew} (hep-homebrew), HEPrpms~\cite{Hahn:2022blh} and 
spack~\cite{spack}, excluding any non-free licensed software parts. In addition, a release on CVMFS for LCG is planned for a future release.


\section{CERNLIB validation}
The validation of the new CERNLIB release is a question of highest priority. The usage of modern compilers allowed fixing some obvious bugs.
The CI system helped to automate the testing.
However, the most important results will come from the
usage of CERNLIB with the preserved software.
 
It is expected that after the adoption of CERNLIB more feedback will be obtained from the users.
 Therefore, the suggested strategy for the validation is to use automation where it is possible and ``to validate vby using it''. This is a community effort.
 
\begin{figure}[htb]
    \begin{center}
    \adjincludegraphics[width=0.45\textwidth, trim={ {0.35\Width} {0.15\Height} {0.20\Width} {0.55\Height}},clip]{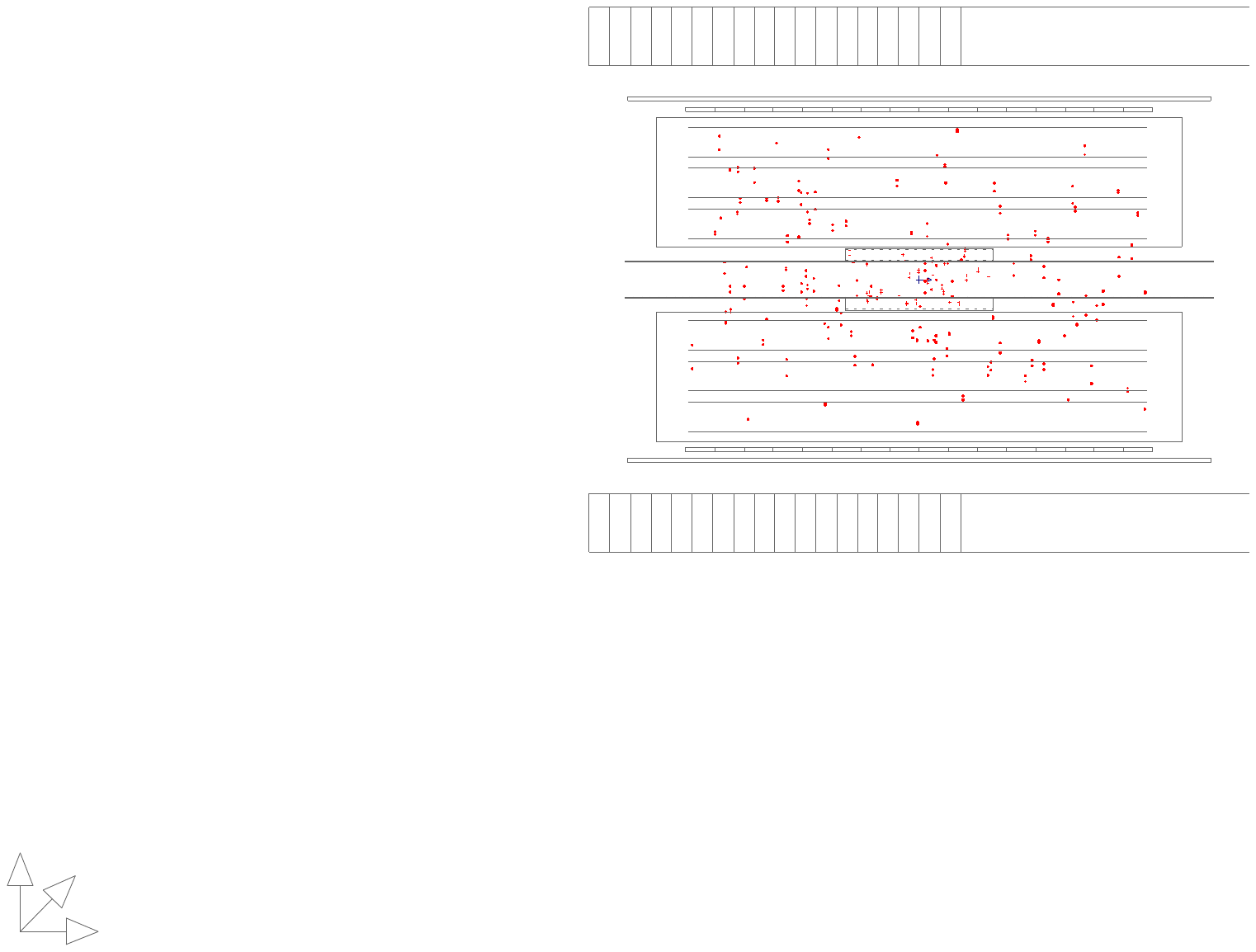}
    \end{center}
    \caption{Example of a JADE event display, produced with CERNLIB 2022.}
    \label{fig:jade}
\end{figure}
\begin{figure}[htb]
  \begin{center}
    \includegraphics[width=0.45\textwidth]{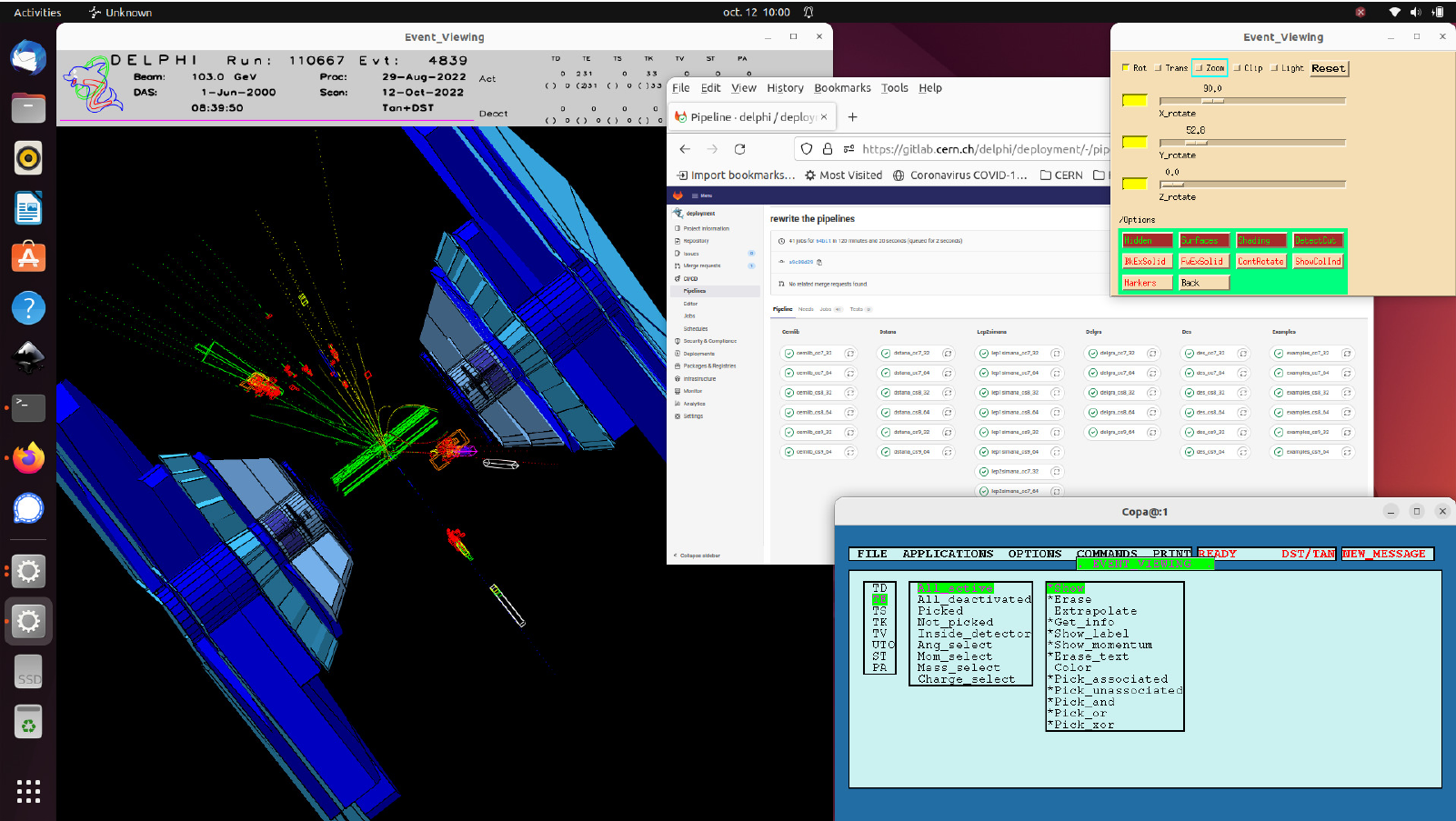}
    \includegraphics[width=0.45\textwidth]{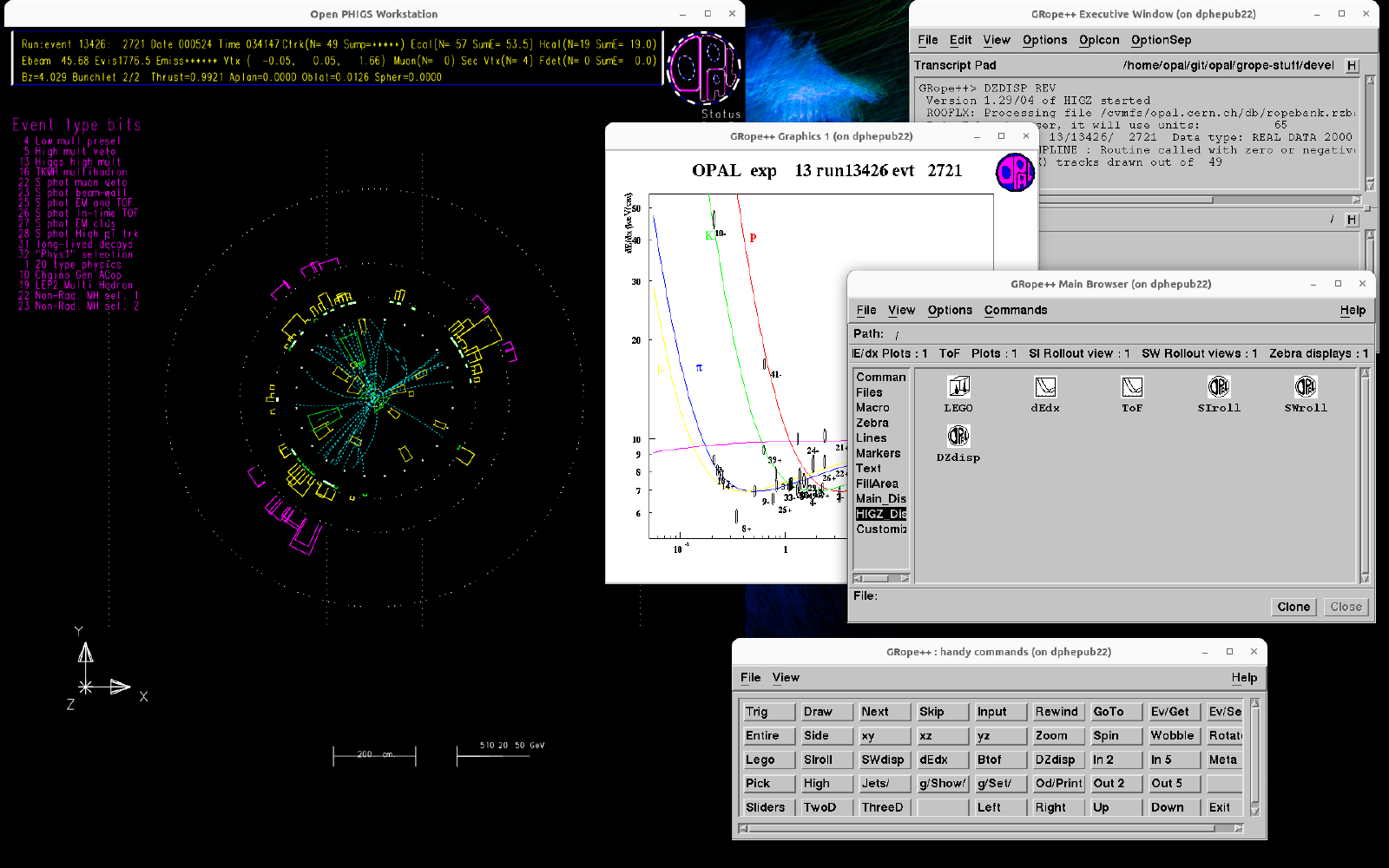}
  \end{center}
  \caption{Event displays of DELPHI (left) and OPAL (right), produced with CERNLIB 2022 in 64-bit using modern compilers and Linux versions.}
  \label{fig:lep}
\end{figure}

\section{Impact on data preservation}

As mentioned already, CERNLIB is a pre-requisite for continued full access to LEP\&JADE data. 
The DELPHI, JADE, and OPAL software stacks were successfully recompiled with the CERNLIB 2022. While validation of software compatibility for JADE is already finished while it is still ongoing in the case of DELPHI and OPAL. No blocking issues have been reported so far.

The availability of 64-bit builds is of particular importance. One reason is that support for 32-bit is vanishing. As an example, Canonical no longer provides 32-bit Motif~\cite{motif} development packages for 32-bit on recent Ubuntu versions. 

The DELPHI and OPAL event displays were originally based on a commercial 3-D software package. In the past years, there was little to no hope to be able to revive these, mainly due to licensing and compatibility issues of the graphics libraries with modern versions of glibc~\cite{glibc}.
This dependency could recently be removed by extending a free, OpenGL-based back-end to the needs of data preservation. This only works properly in 64-bit mode though, and thus requires a full 64-bit CERNLIB and experiment software stack. Good progress has been made in this direction. DELPHI already released a new version of it, and for OPAL a proof of concept exists that works on Ubuntu 22 and CentOS 8.
Fig.~\ref{fig:lep} shows screenshots of the two event displays.

Modernisation of CERNLIB enables the adoption of modern tools, e.g.\ automatic building and testing for the whole software stacks depending on CERNLIB. In the case of DELPHI, a CI has been implemented which allows rebuilding of the whole stack for a variety of operating system versions on changes, including basic checks for completeness of the stack. The resulting artifacts can directly be used for deployment. A similar approach is planned for OPAL in the near future.

\section{Further Plans}
While initial testing looks very promising, and no major issues have been identified so far, more evaluations are needed to confirm that the new code base works and can be used to reproduce old results from the different experiments.
This implies further advertising of it's availability and timely bug fixing as reports come in.
Maintenance will have to continue, on a voluntary basis, to adapt the code and build infrastructure to upcoming changes. This includes upstream software evolution, for example in terms of compilers, and also changes in the computing architecture, in particular new generations of CPUs.

\section{Conclusions}
The presented approach to revive CERNLIB combines experiences from various experiments that depend on it with modern tooling, allowing to automate the creation of the libraries and tools. This makes it easy to be ported to additional platforms while maintaining backward compatibility, making it fit for data preservation for the coming 10 years and beyond.

\section*{Acknowledgements}
We are grateful to the authors of patches to CERNLIB that were created since the official support of CERNLIB was terminated and to the Fedora and Debian projects that maintained the builds of CERNLIB in their repositories.

We are grateful to the authors of the Pythia6, ARIANDE, LEPTO~\cite{Ingelman:1996mq}, JETSET~\cite{Sjostrand:1993yb} and fritiof~\cite{Nilsson-Almqvist:1986ast} packages who agreed to re-license their codes under GPL license for CERNLIB.

\section*{References}
\bibliography{CL}
\end{document}